# Effects of a Conducting Sphere Moving Through a Gradient Magnetic Field

Adom Giffin, Mikhail Shneider, Chiranjeev S. Kalra and Richard B. Miles
*Applied Physics Group*
*Department of Mechanical and Aerospace Engineering*
*Princeton University*
*Princeton, NJ 08540,USA*

*and*

T. L. Ames
*Sandia National Laboratory, 1515 Eubank SE*
*Albuquerque, NM 87123, USA*

**We examine several conducting spheres moving through a magnetic field gradient. An analytical approximation is derived and an experiment is conducted to verify the analytical solution. The experiment is simulated as well to produce a numerical result. Both the low and high magnetic Reynolds number regimes are studied. Deformation of the sphere is noted in the high Reynolds number case. It is suggested that this deformation effect could be useful for designing or enhancing present protection systems against space debris.**

## I. Introduction

As a conducting body moves through a magnetic field, it normally experiences a force to oppose the motion due to eddy or Foucault currents[1]. These are created from the EMF that is produced by the magnetic flux change. However, if the magnetic field is uniform, the EMF that is produced initially is canceled by the electric field that is also produced due to charge separation. This is not always well understood[2].

In this paper, we examine the effects of a conducting sphere moving through a *gradient* magnetic field. Although the description is simple, the physics behind the solutions to these effects are not only understudied, but can be very complicated as well. Particularly when the conductor is moving very fast (non-relativistic) or better stated, when the magnetic Reynolds number[1] is large. We attempt to discuss this problem by first examining a simple situation (low velocity) where we can describe the analytical solution, then move to a more complicated scenario (high velocity). Although there are numerical solutions to these cases, those solutions often fail to show the relationships between properties of the system. In fact as J. B, Jackson[3] says, "Faraday's law of induction is somewhat tricky to apply to moving circuits...".

For the low Reynolds number case, we conduct both an experiment as well as solve the problem by way of an analytical approximation. For the experiment, we have a superconducting magnet based on a Helmholtz coil design that is fixed at 4 Tesla at the center. Various solid metallic spheres are dropped into the magnetic field and the fall time from release to the arrival at the center of the coil pair is recorded using laser beams, as shown diagrammatically in figure 2. The interaction of the magnetic field with these spherical objects leads to the development of eddy currents that produce an induced magnetic field which then generates a force opposing the gravitational force, thus reducing the fall velocity and increasing the fall time.

For the analytical model approximation, we use a piecewise linear approximation for the magnetic field. The most important feature of the field is the gradient and our linear approximation accurately captures that slope. The relevant quantity is $B_{max}^2/L^2$, the maximum magnetic field divided by the linear dimension over which that field changes from 0 to the maximum value. The analytical approximation and the experimental results have very close agreement.

For the high velocity case, we use a commercial software package, MagNet by Infolytica Corporation. In this case, the Helmholtz coil configuration and spheres are modeled with the software. A low velocity simulation shows good agreement with both the analytical approximation and the actual experiment. Following this, the high velocity case is simulated. While the velocity decrease is only slight in this case, there are significant differential forces built up over the sphere. These forces seem to flatten and extrude the sphere into a prolate spheroid.

One possible application for this work would be for protection from space debris. The current configuration of the ISS shielding uses a "bumper" and "catcher" setup[4]. The idea is that the debris will hit the bumper (aluminum), melt due to impact and spread out (due to design) across the catcher, thus distributing the force over a greater area. If this system also included a magnetic field, the forces on the mass could be further spread out and slowed. This would result in the need for less shielding.

## II. Low Magnetic Reynolds Number Case

In this case we will examine a sphere moving through a magnetic field gradient at low velocity or with a low magnetic Reynolds number, $R_m$. We will define the magnetic Reynolds number as,

$$R_m = \mu\sigma v l, \tag{1}$$

where $\mu$ is the magnetic permeability, $\sigma$ is the conductivity, $v$ is the velocity and $l$ is the characteristic length which in our case will be the radius of the sphere.

### A. Analytical Approximation

To begin, we wish to calculate the net force acting against the sphere due to a gradient magnetic field. We will examine the simple case of the sphere traveling transversely through a field that is only changing linearly in the direction of motion. For the analytical approximation, we will assume that the velocity is *constant*. The purpose of this is to reduce complexity of the calculations. As can be seen in figure 1, we expect that there will be a net force that opposes the sphere due to the gradient.

Since we are in the low Reynolds number regime, we can neglect any perturbations to the magnetic field. With this and our previous assumptions, we solve the appropriate Maxwell equations to arrive at an equation for the opposing force,

$$F_{-v} = \frac{2\pi\sigma}{15}\frac{B_{max}^2}{L^2} v r^5, \tag{2}$$

where $B_{max}$ is the maximum point of the field, $L$ is the length of the gradient, $r$ is the radius, $v$ is the constant velocity and $\sigma$ is the conductivity. Even though we initially assume a constant velocity, we relax this assumption in order to determine the motion experienced in the field by solving the equation of motion using this force.

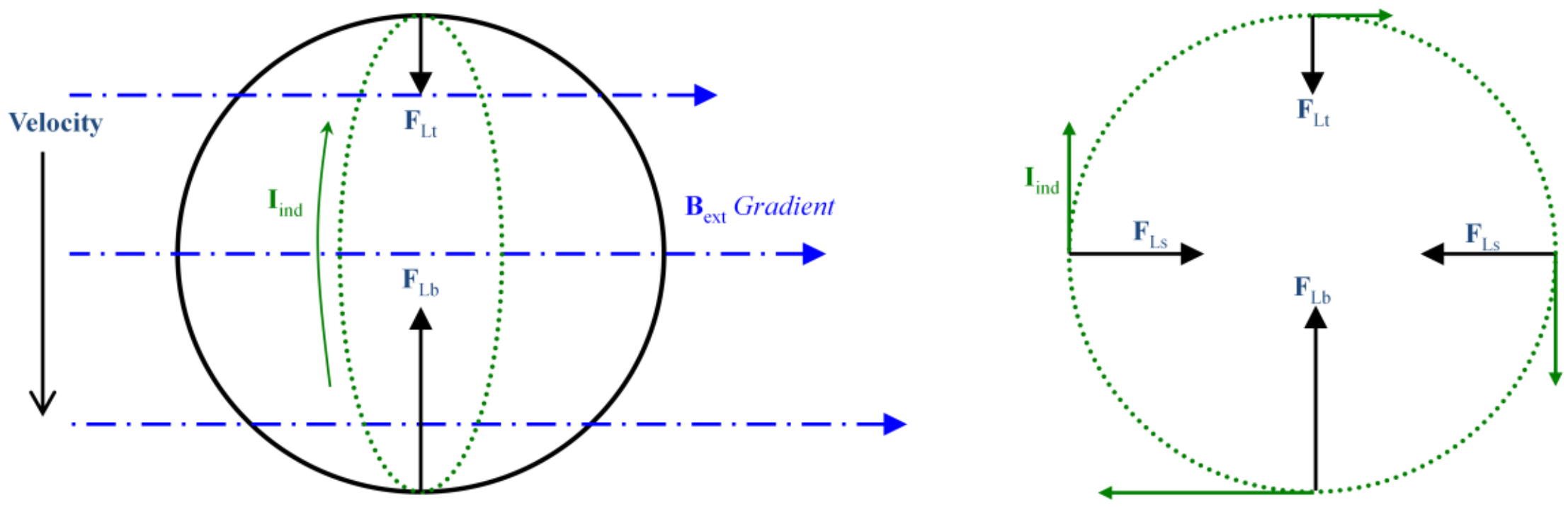

**Figure 1**. *Induced current and induced forces associated with a sphere moving through a positive magnetic field gradient. The left diagram is a view of the sphere seen from a position orthogonal to the magnetic field lines and the right diagram shows a cross section of the sphere seen from a position along the magnetic field lines. Note that because the gradient is positive, the induced magnetic field is in the opposite direction of the external field.*

### B. Experimental Results

To verify the analytical approximation we perform an experiment using a magnetic consisting of a pair of superconducting Helmholtz coils capable of operating at up to 6 Tesla. Access to the high magnetic field at the center of the magnet is by three orthogonal passages, including a vertical passage through which spheres are dropped and a horizontal passage through which a laser beam is passed for timing arrival of the spheres at the center.

For the experiment, we use three spheres: an aluminum sphere (Al 2011) with a 1" diameter, an aluminum sphere (Al 2017) with a ¾" diameter and a copper sphere (Cu 145) with a ½" diameter. The spheres were released above the magnet and the fall time to the center of the magnet was recorded using a laser sheet, as shown diagrammatically in figure 2.

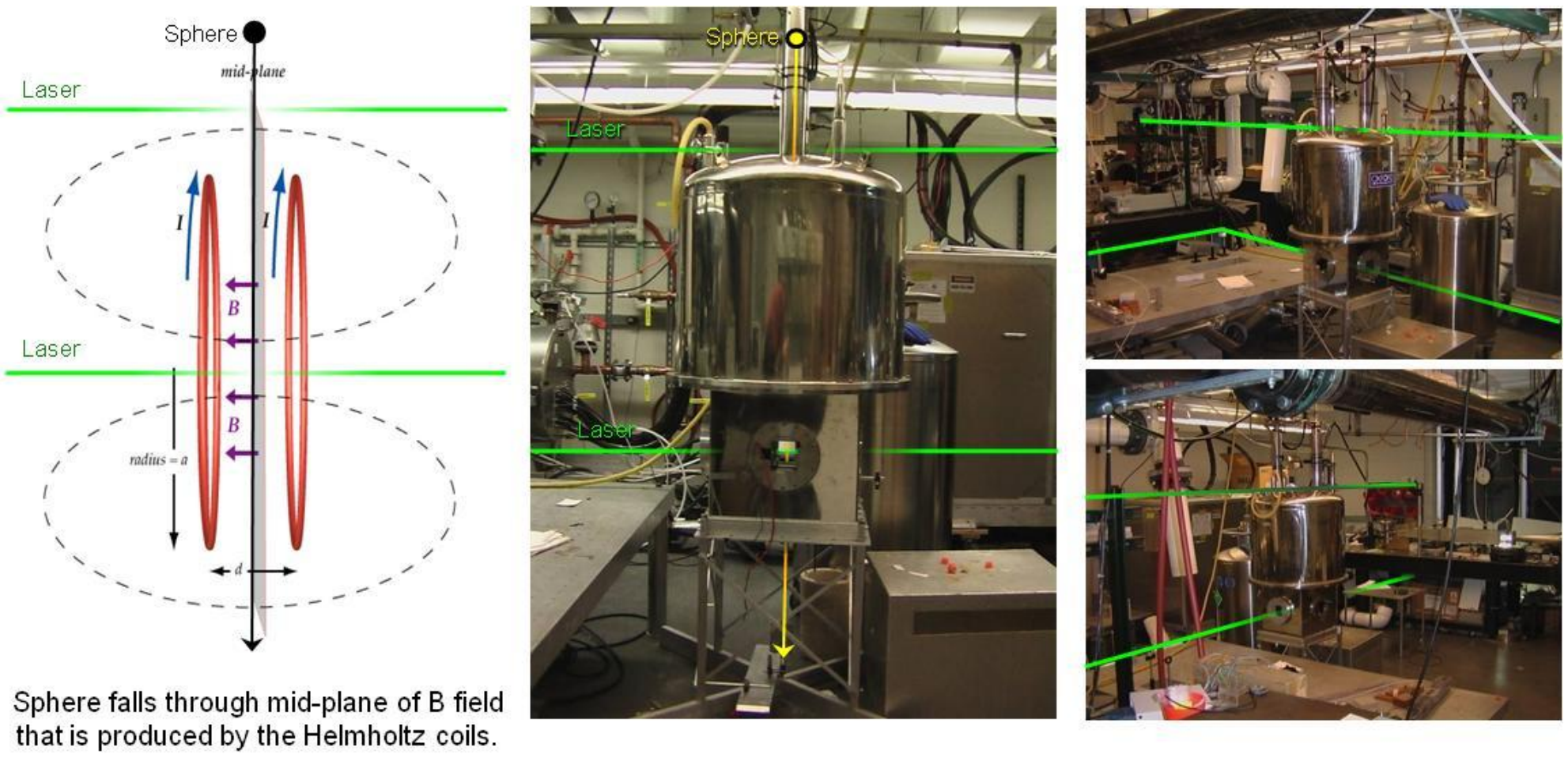

**Figure 2.** *Geometry of the Helmholtz magnetic coils and photographs of the laboratory set up. Spheres are dropped through a vertical hole (yellow line shows the fall trajectory) and pass into the high field region at the center of the magnet coils. The fall time is recorded by a pair of laser beams as shown in green.*

As the spheres move through the magnetic field, eddy currents are produced with creates a small induced magnetic field that opposes the external field. This field, along with the eddy currents produce a force that opposes the motion of the sphere, thus increasing the fall time. To compare this with the analytical approximation, we solve for the axial component (x component) of the magnetic field, along the transverse or radial direction through the center between the coil pair as indicated in figure 3 using the specifications of the magnet.

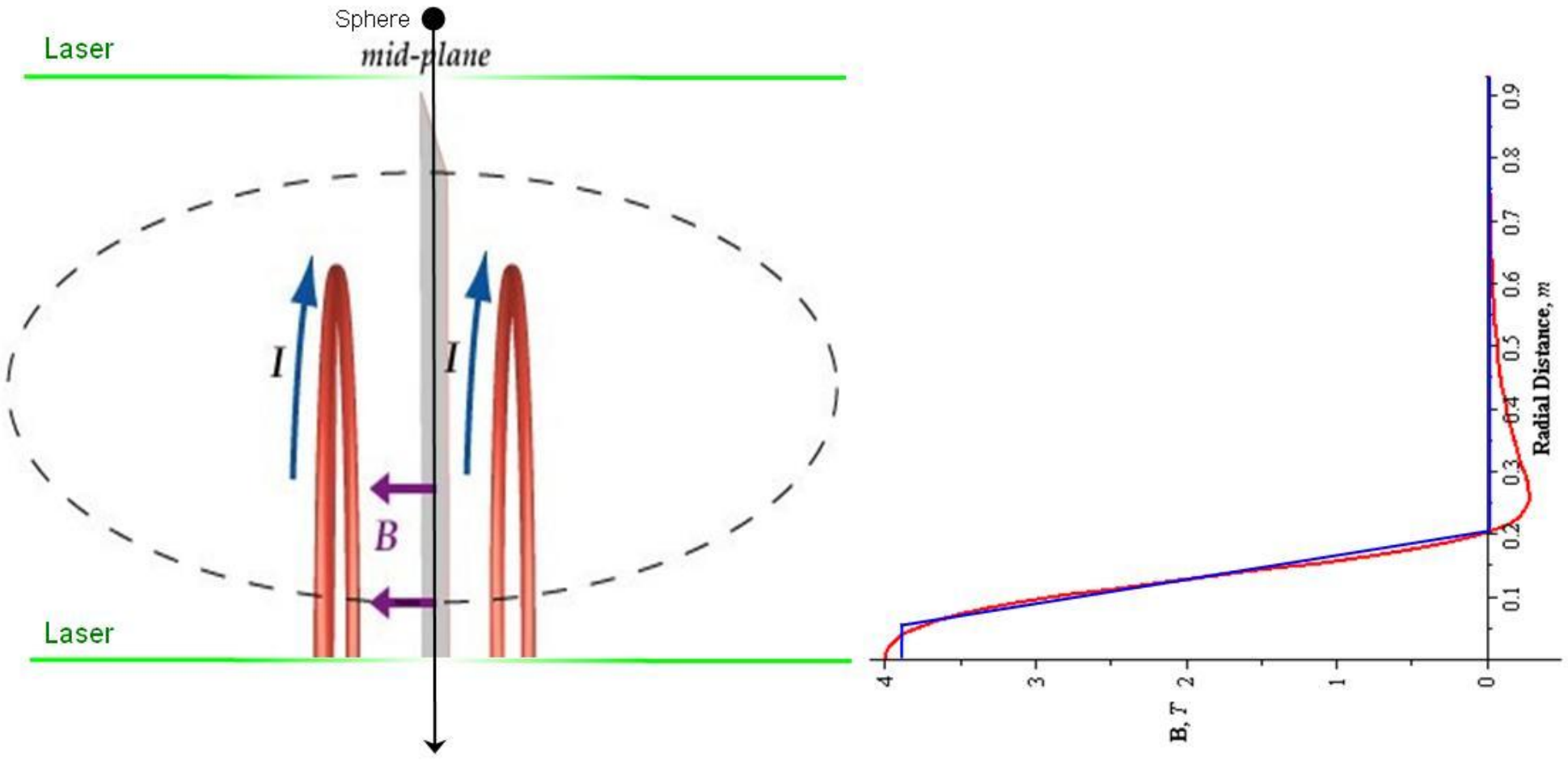

**Figure 3.** *Modeled and actual magnetic field gradient. The actual gradient is calculated from the specifications given for the magnet. The modeled field is a piecewise linear approximation.*

The analytical model uses a piecewise linear approximation shown in blue. The most important feature of the field is the gradient and the linear approximation accurately captures that slope. The relevant quantity is $B_{max}^2/L^2$ where the maximum magnetic field is divided by the linear dimension over which that field changes from 0 to the maximum value. Accordingly, the values for the spheres are also used in the analytical solution to determine an approximate fall time.

Figure 4 shows the results of the drop tests and the analytical predictions. Both aluminum and copper alloy spheres were tested. The significant difference in the fall times associated with the radius of the sphere as predicted by the analytical approximation in Eq. (2). Note that the fall time for the solid 1" aluminum sphere is almost three times that of free fall. The analytical model captures the fall times to within 4% of the measured values.

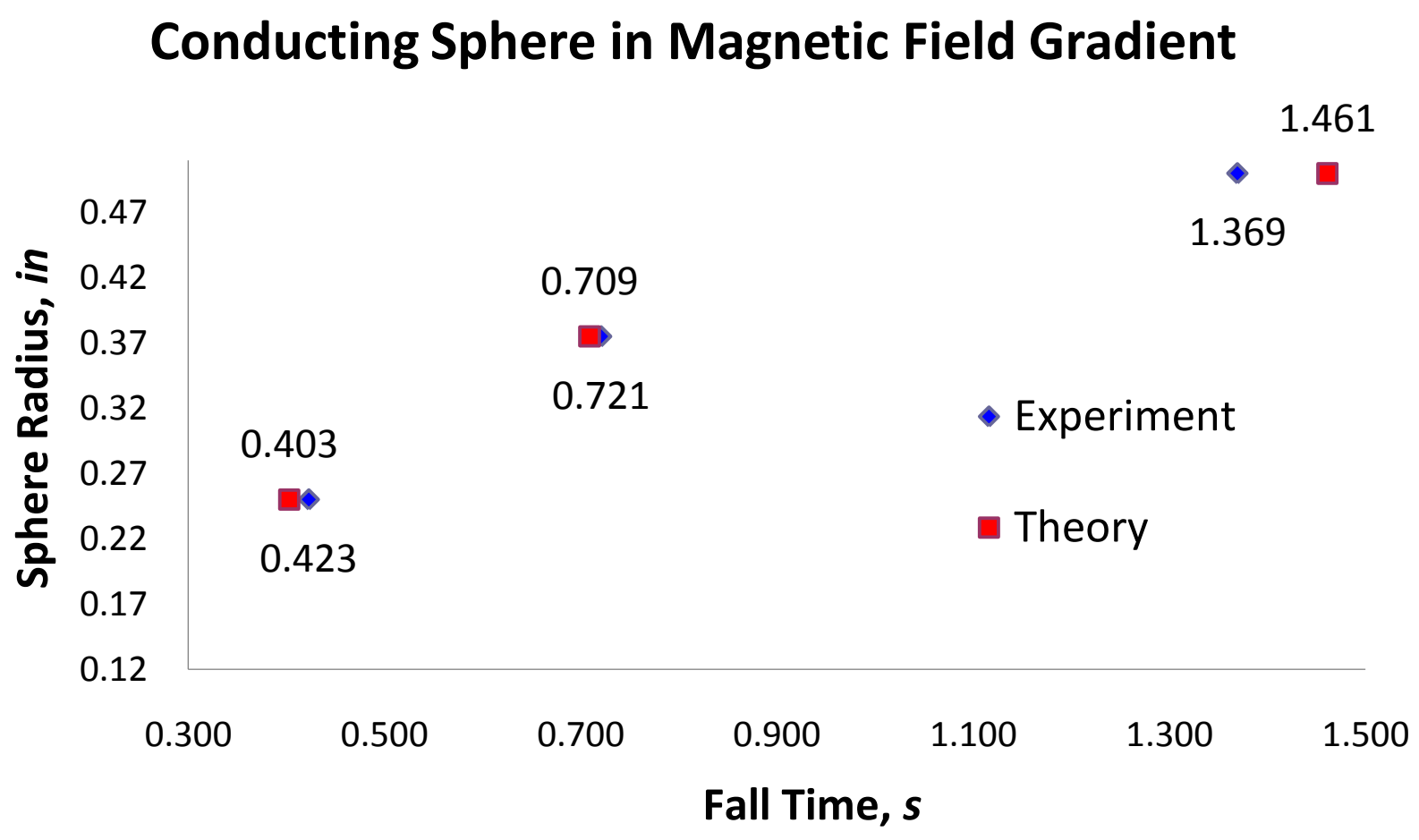

**Figure 4.** *Results for drop tests for copper and aluminum solid spheres. The diagram shows the recorded and predicted fall times. Note that the aluminum alloys had conductivities that were significantly lower than the copper. Error bars are represented by the size of the markers.*

## C. Numerical Solution

In addition to the analytical and experimental solutions, we also use a modified commercial code called MagNet by Infolytica to simulate the experiment. This is done so that we can use our previous results as a benchmark for the numerical solution. This is needed as we will rely more heavily on the numerical solutions for the high Reynolds number cases. As can be seen in figure 5, the simulated results correspond very well to the analytical approximation.

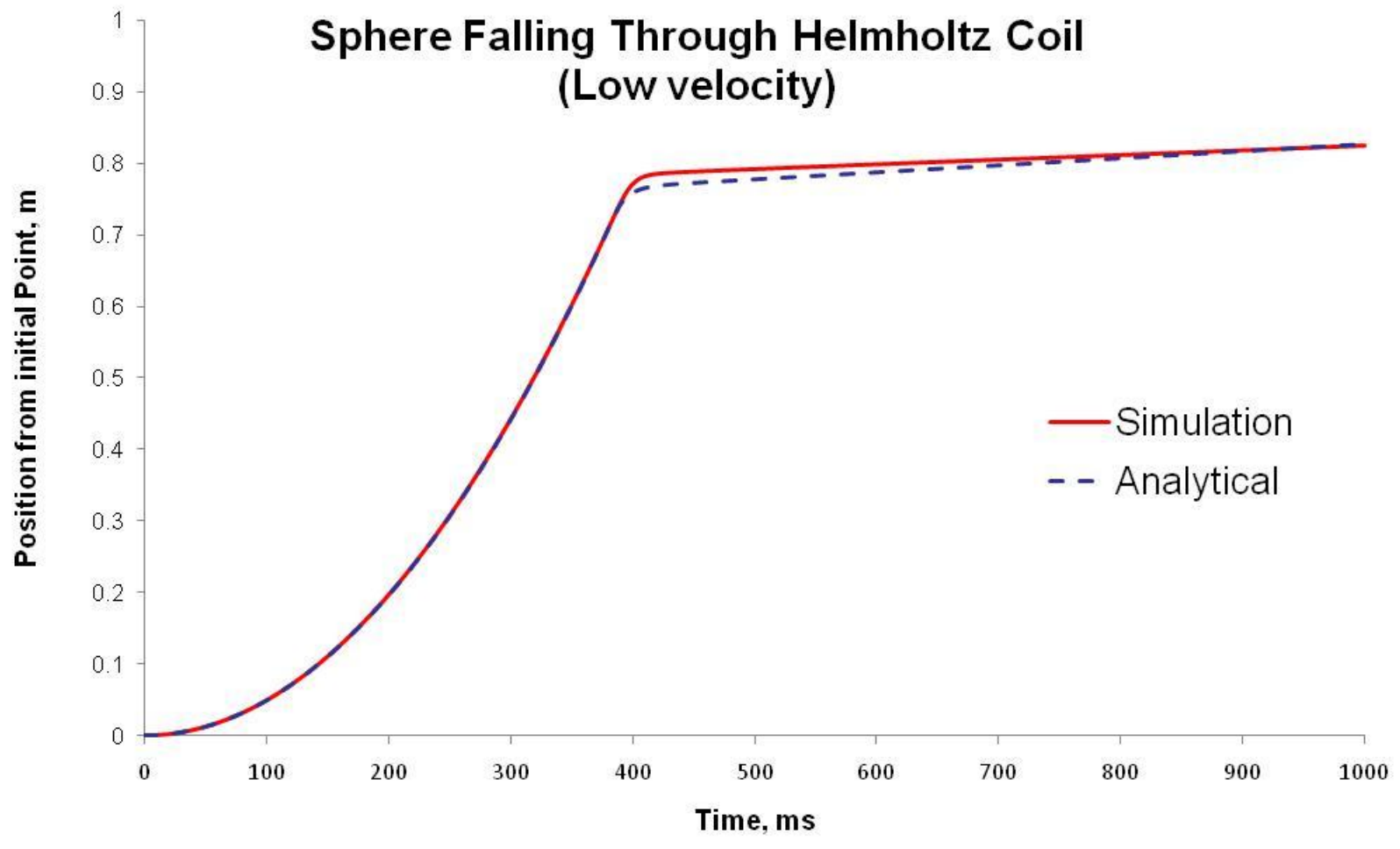

**Figure 5.** *Comparison of analytical approximation and simulated numerical results using commercial code.*

# III. High magnetic Reynolds Number Case

In the high magnetic Reynolds number case, we cannot rely on our analytical approximation because the induced magnetic field is much larger. In this case we are examining objects with velocities that exceed 1 km/s.

## A. MagNet Simulation

We simulate a sphere moving in the high Reynolds number regime. In the case of a copper sphere moving at 1 km/s, the reduction in velocity is minimal over the distance of the gradient, even though there is a large deceleration as shown in figure 6. This is due to low momentum exchange over the time of the travel.

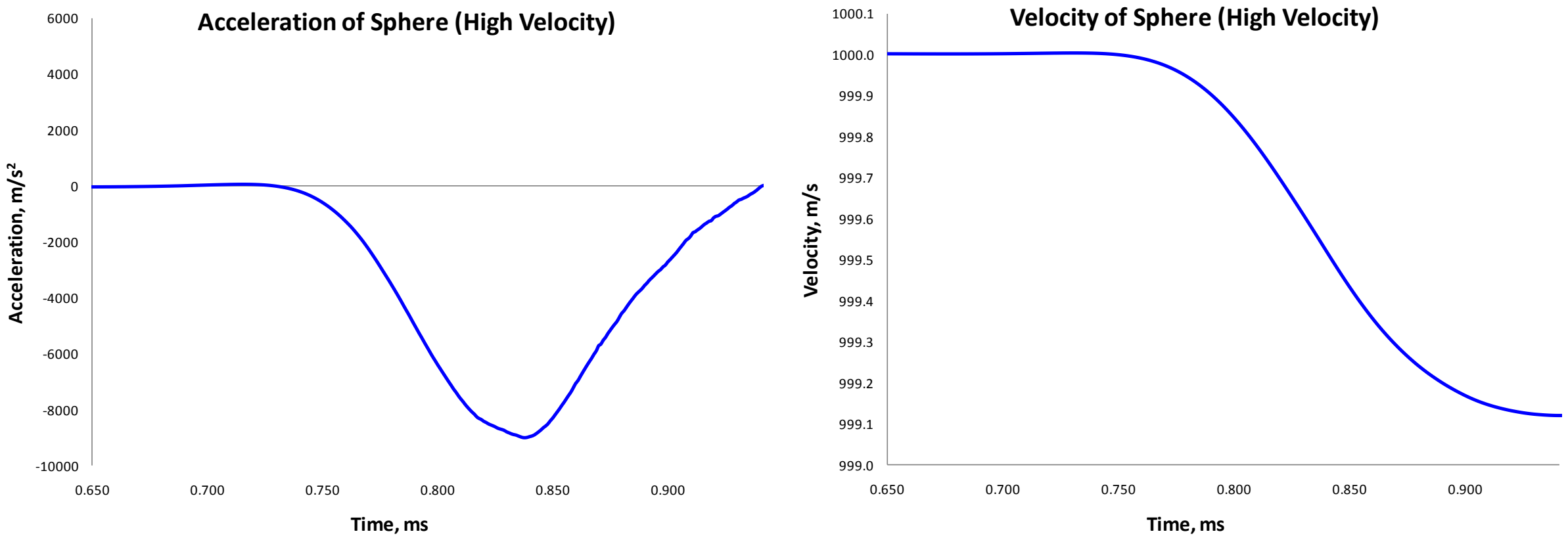

**Figure 6.** *These two graphs show the changes in the acceleration and velocity as a conducting sphere moves through the magnetic field gradient (0 to 4 Tesla over ~20cm) at low velocity (under the influence of gravity for 1 meter).*

However, the deceleration and therefore the force is very large. This implies that while at this field level and velocity we cannot expect the object to slow down significantly, we may expect it to *deform*. This can be seen in figure 7 which shows a 3-D surface picture of the induced force on a solid copper sphere with a radius of 12.7 mm (1" diameter) at 1 km/s. The magnetic field extends from 0 T to 4 T over a span of ~0.2 m in the -y direction. As the sphere travels through the gradient, the color and arrows indicate the strength (red = strongest) and direction (approximately radial) of the pressure on the sphere. The largest forces are at the bottom of the sphere which has the effect of slowing down the sphere. The forces become very large as the velocity or gradient increases due to the increased strength of the eddy currents. Therefore, the key aspect to note is the differential of the forces acting on the sphere. These forces will flatten and extrude the sphere into a prolate spheroid shape. Thus the pressure of the impact of the sphere against another object will be diminished.

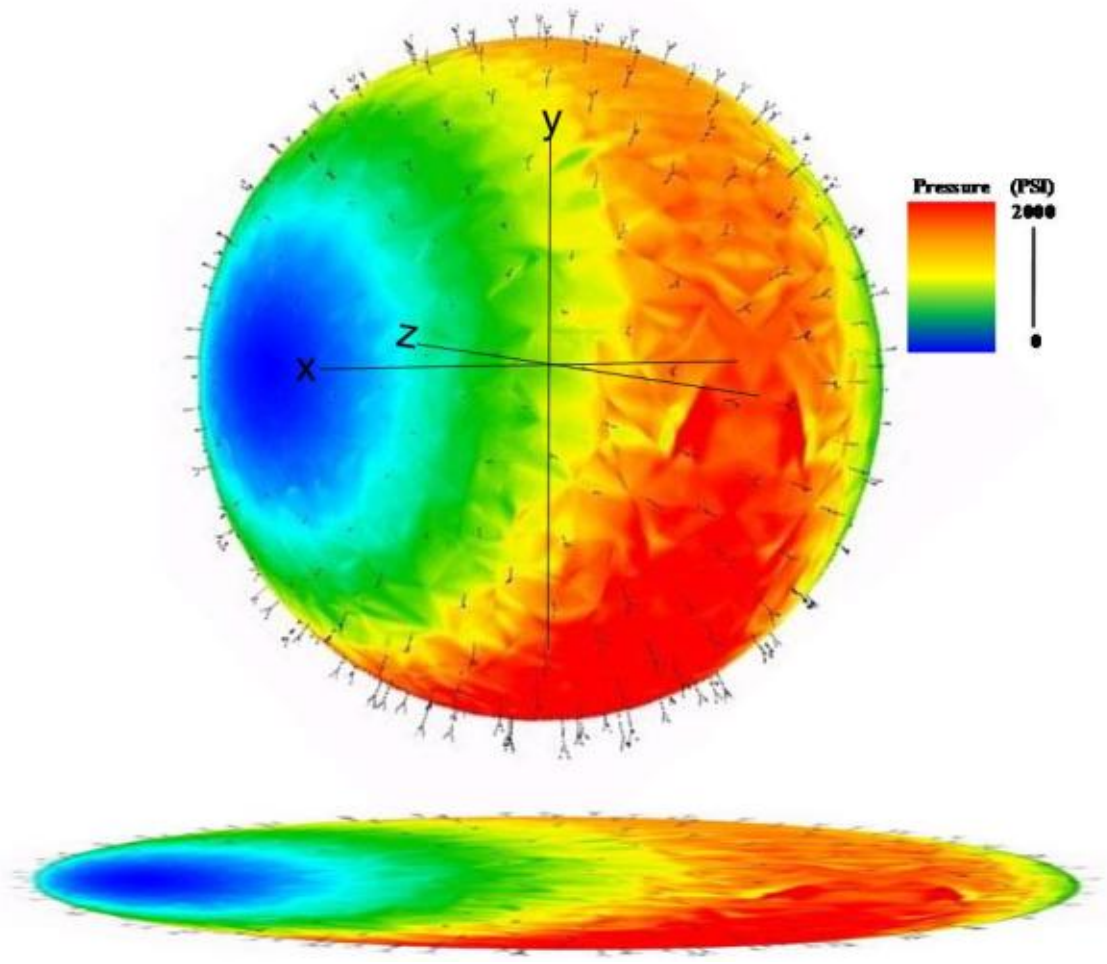

**Figure 7.** *The top picture is a 3-D surface picture of a sphere moving through a magnetic field gradient. The sphere is moving in the –y direction under the influence of gravity. As the sphere travels through the gradient, the color and arrows indicate the strength (red = strongest) and direction (~radial) of the pressure on the sphere. The bottom picture is a 3-D representation of the sphere above as forces flatten and extrude the sphere into a prolate spheroid shape.*

## B. Alegra Simulations

One of the advantages of the using the modified MagNet software is that it can model the well known skin effect[2] very well. The drawback is that the objects in the software cannot be deformed. Therefore we needed to use a more robust code. For this we used the Alegra code at Sandia National Laboratory. After confirming that the Alegra code results agreed with our previous results, we simulated the sphere moving at various speeds and magnetic field strengths in order to examine the deformation of the sphere. When the magnetic field was 4 T, there was very little deformation of the solid sphere. However, at 40 T the sphere was shown to deform significantly as seen in figure 8. This also shows that the greatest deformation does not occur at the highest velocity due to the finite interaction time in the field.

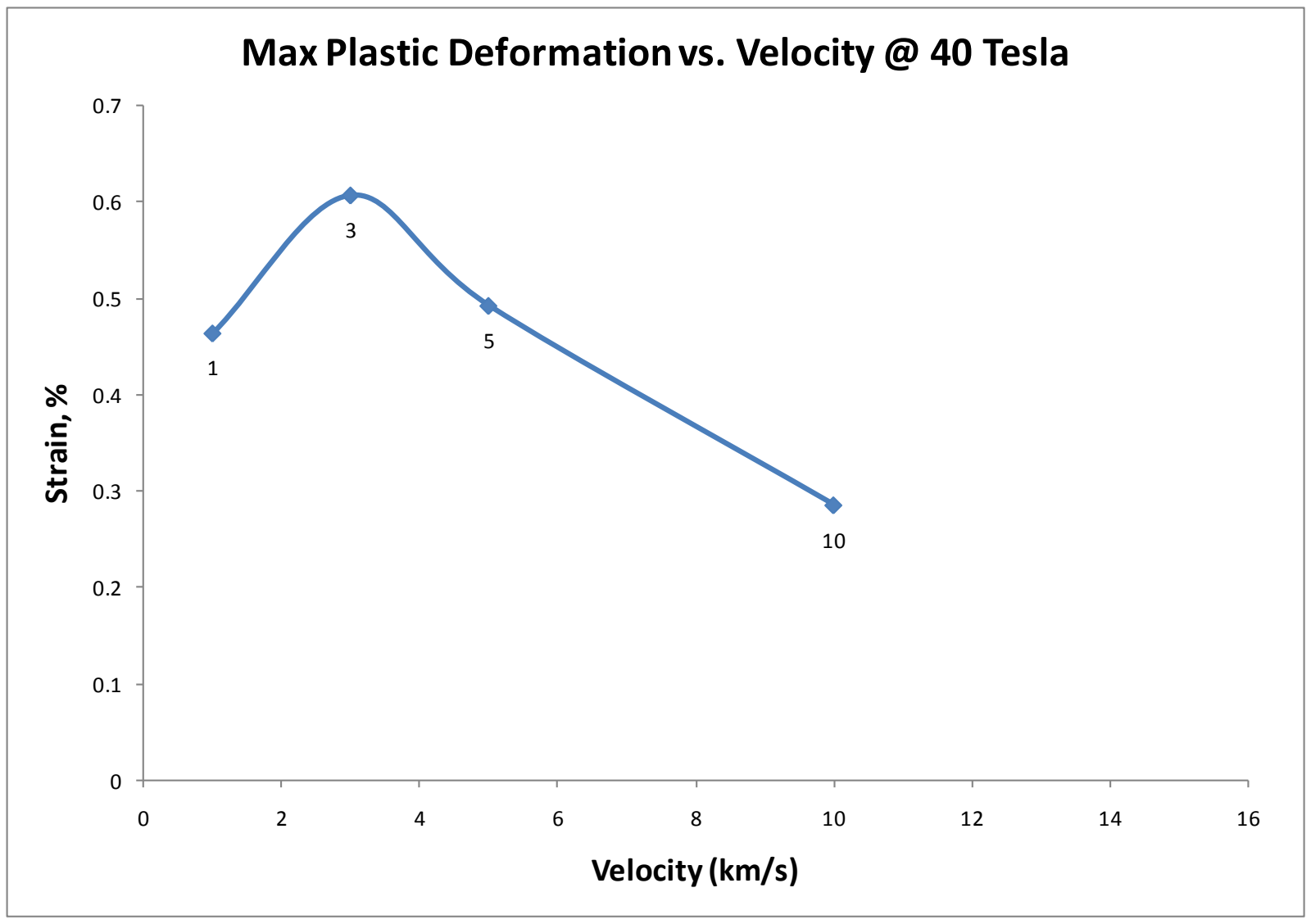

**Figure 8.** *This figure shows the percent strain of a solid copper sphere moving at various velocities in a gradient with a maximum of 40 Tesla. Notice the peak strain is not during peak velocity.*

## C. High Reynolds Number Approximation

Although this regime is characterized by non-linear currents and a highly perturbed magnetic field, we can still make a qualitative estimation, if not a roughly quantitative approximation for the force under these conditions using Eq (2). In this case, we can assume that the current is mostly confined in the skin layer of the sphere. Therefore, if we calculate the force that is due to the shell of the sphere that has a thickness of the skin layer, we should see reasonable results which are shown in table 1.

**Table 1**

| Mag Reynold's # | Simulation, m/s2 | Approximation, m/s2 | Percent Error |
|---|---|---|---|
| 350 (1000 m/s) | 12444 | 14060 | 13.0% |
| 920 (2631 m/s) | 15400 | 23338 | 51.5% |
| 1750 (5000 m/s) | 16325 | 32585 | 99.6% |

As can be easily seen, the change in the magnetic Reynolds number is almost directly equivalent to change in the percent error. We can adjust for this with an appropriate adjustment to the skin depth,

$$\delta = \frac{\mathrm{r}}{\sqrt{\mathrm{R_m}}} \frac{\mathrm{e}^{-.0004 \cdot \mathrm{R_m}}}{\sqrt{\pi}}, \tag{3}$$

where the left fraction is the diffusion depth definition[5] rewritten using the Reynolds number and the right fraction is our empirical adjustment. Note: Classically the skin depth is equal to $\sqrt{2/\mu\sigma\omega}$ where $\omega$ is the angular frequency of the signal. However, this is for the special case of a sinusoidal signal.

## IV. Protection from Orbital Debris

The current configuration of the International Space Station (ISS) shielding uses a "bumper" and "catcher" setup as seen in figure 9. The idea is that the debris will hit the "bumper" (aluminum), melting due to impact and spread out (due to design) across the "catcher", thus distributing the force over a greater area. Since, as previously stated, in the high magnetic Reynolds regime the currents are produced in the outer surface (skin), the effect of the magnetic field would be concentrated on the aluminum. This would cause the aluminum to spread out and induce a large force on the incoming debris so that it spreads out as well. Therefore, if this system also included a magnetic field as shown, the forces on the mass could potentially be further spread out and slowed. This may result in the need for less shielding. Additionally, the current protection configuration is weakest when the debris is larger than 1" or is not moving fast enough to melt on impact. Our implementation could also help in this regard for metallic objects. The force is proportional to radius, $r^5$ so the bigger the object, the greater the effect. Also, the slower the debris was coming in, the more time would exist for deformation. We suggest that this is feasible using currently produced high temperature superconducting materials.

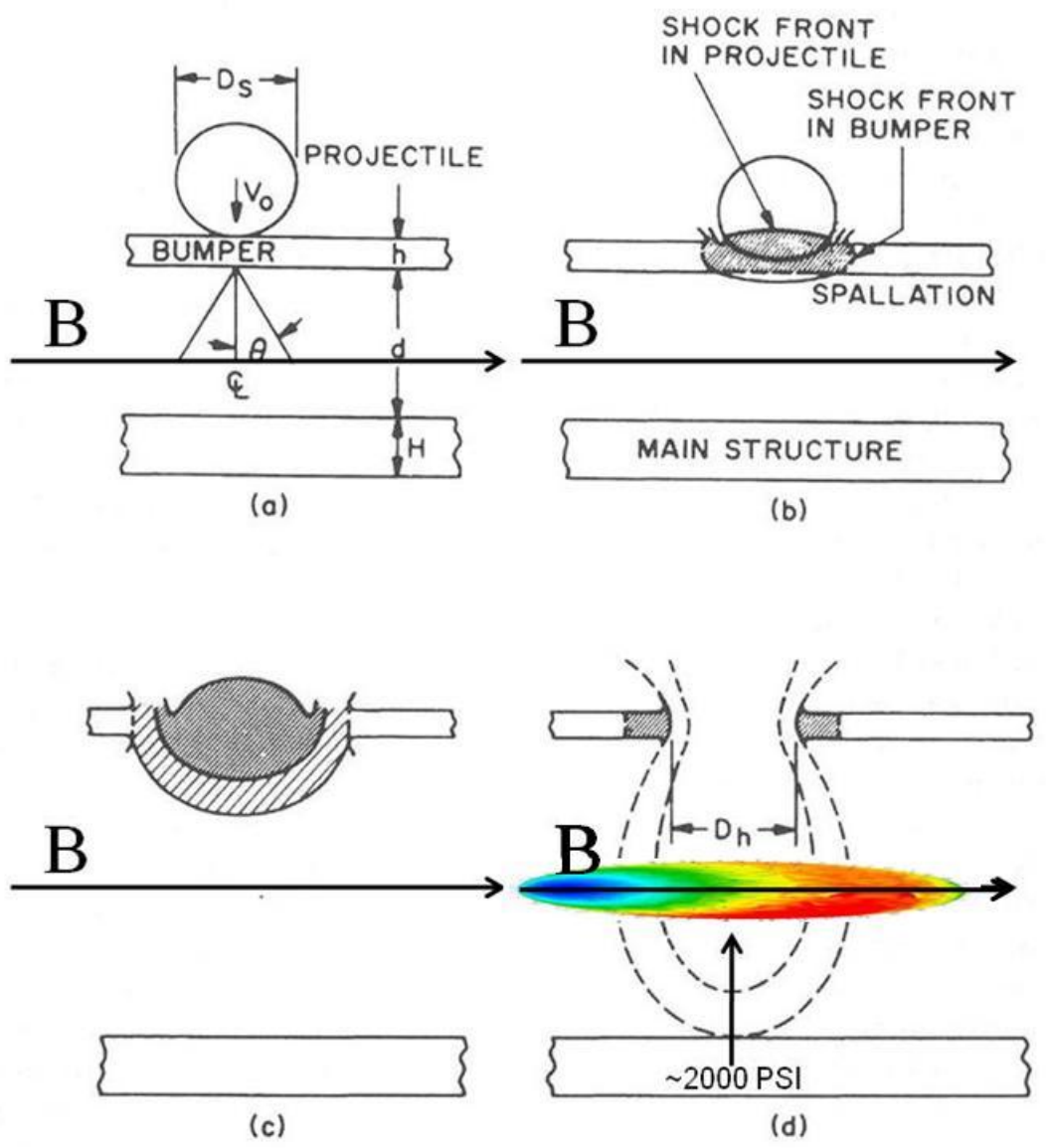

**Figure 8.** *As the debris (sphere) hits the "bumper", melting due to impact, the material is spread out (due to design) across the catcher[4]. This distributes the force over a greater area. If this system also included a magnetic field, the forces on the mass could be further spread out and slowed. Note that the 2000 PSI in the diagram refers to a 4 T field. A 40 T field would produce 100 times more pressure.*

## V. Conclusions

An analytical approximation was derived for a sphere moving at low velocity through a gradient magnetic field. This was confirmed by way of an experiment using a superconducting magnet and a variety of conducting spheres. This approximation give us a reasonable understanding of the variables and their magnitudes that contribute to the forces induced. In addition, the change in temperature was calculated and not found to be significant.

For the high magnetic Reynolds number case, we see significant deformation of the sphere. Therefore, a magnetic field could be used to deform a projectile moving at a high rate of speed so as to spread the force of the impact out over a larger area. While 40 T is an extremely high field to achieve practically, note that the simulations were performed using a room temperature solid sphere. The point here is that as the sphere collides with the bumper, as in the ISS configuration, the sphere will tend to melt due to impact friction and compression. This would greatly reduce the yield strength of the sphere and thus would deform at much lower field strengths. An additional strength is that the magnetic field does not need to come on right at impact. With a superconducting magnetic, the field can be energized ahead of time and run in a persistent mode.

## Acknowledgments

This work was supported by DARPA DSO under Dr. Judah Goldwasser. Sandia National Laboratories is a multi-program laboratory operated by Sandia Corporation, a wholly owned subsidiary of Lockheed Martin Corporation, for the U.S. Department of Energy's National Nuclear Security Administration under contract DE-AC04-94AL85000.